\documentclass[pra,twocolumn,showpacs,amsmath,amssymb,floatfix,letterpaper]{revtex4}

\usepackage{graphicx}
\usepackage{dcolumn}
\usepackage{bm}

\begin{document}


\title{Two-photon photo-ionization of the Ca $4s3d \; ^1D_2$
level in an optical dipole trap}

 \author{J. E. Daily}
 \author{R. Gommers}
 \altaffiliation[Also at ]{Physics Department, University College
London, Gower Street, London, WC1E 6BT, UK}
 \author{E. A. Cummings}
 \author{D. S. Durfee}
 \author{S. D. Bergeson}
 \email{scott.bergeson@byu.edu}
 \affiliation{
 Brigham Young University,
 Department of Physics and Astronomy,
 Provo, UT 84602}

\date{\today}

\begin{abstract}
We report an optical dipole trap for calcium. The trap is created
by focusing a 488 nm argon-ion laser beam into a calcium
magneto-optical trap.  The argon-ion laser photo-ionizes atoms in
the trap because of a near-resonance with the $4s4f \; ^1F_3$
level. By measuring the dipole trap decay rate as a function of
argon-ion laser intensity, we determine the $^1F_3$
photo-ionization cross section at our wavelength to be
approximately 230 Mb.
\end{abstract}


\maketitle

\section{Introduction}

Laser-cooling experiments have expanded in recent years to group
II metals.  This is partially due to a growing interest in
optical frequency standards \cite{ruschewitz98,oates99,riehle99}.
Calcium, strontium, and ytterbium all have narrow resonances from
the ground state at relatively convenient laser wavelengths.  The
major isotopes of these elements have no angular momentum in the
ground state, making the narrow ``clock'' transition frequencies
less sensitive to external fields.  Other experiments, such as
metastable collision studies, photo-associative spectroscopy and
quasi-molecule formation, Bose-Einstein condensation in simple
atomic systems, and ultracold plasma investigations also
contribute to the growing interest in laser-cooling
alkaline-earth metals.

The atomic density in dipole traps can be much higher than in
magneto-optical traps.  For some experiments, such as
Bose-Einstein condensation, photo-association, cold plasmas, and
collision studies, higher densities can be helpful. Dipole traps
for alkaline-earth atoms may also improve the performance of
atomic clock experiments.  For example, atoms can be held in a
dipole trap generated by a laser at the so-called ``magic
wavelength'' where the ac Stark shift is exactly equal for both
atomic levels in the clock transition
\cite{takamoto03,degenhardt04}.  This would make it possible to
use trapped neutral atoms for the clock, increasing the maximum
interrogation time and therefore increasing the accuracy of the
clock.

Only a few experiments have explored optical dipole traps for
alkaline-earth atoms.  The absence of angular momentum in the
ground state prevents sub-Doppler cooling using resonance
transitions \cite{ye03}, complicating dipole trap loading.
However, advanced cooling techniques can reduce the atomic
temperature to a few microKelvin, and dipole traps in Sr
\cite{katori99,katori03} and Yb \cite{takasu03a,takasu03b} have
been reported in the literature.  We are also aware of a
ground-state calcium dipole trap reported in a Ph.D. thesis
\cite{degenhardt04a}.

In this paper we report an optical dipole trap for calcium. Our
trap captures $T\sim 1$ mK atoms in the excited $4s3d \; ^1D_2$
metastable state. Because the atomic temperature is relatively
high, the dipole trap operates in the regime where the light-shift
is several times larger than the natural atomic transition
linewidth \cite{heinzen93}. The high intensities required to
capture these atoms is also high enough to photo-ionize them. By
measuring the photo-ionization production rate and the trap
lifetime as a function of dipole trap laser intensity, we
determine the effective $^1D_2$ lifetime in our system and
photon-ionization cross section.

\section{Magneto-optical trap}

\begin{figure}
\includegraphics[angle=270,width=2.6in]{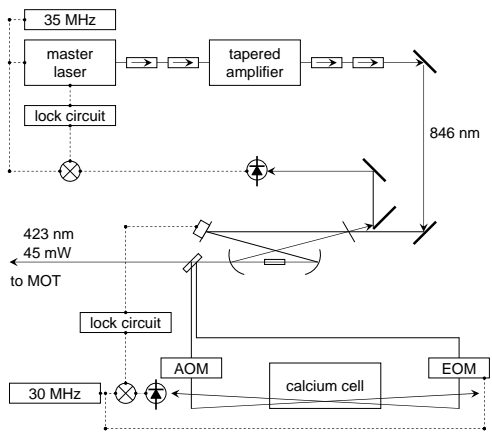}
\caption{\label{fig:exptSketch} A schematic drawing of the MOT
laser system and frequency stabilization electronics used in
these experiments.}
\end{figure}

The calcium MOT is formed by three pairs of counter-propagating
laser beams that intersect at right angles in the center of a
magnetic quadrupole field \cite{raab87}.  The 423 nm laser light
required for the calcium MOT is generated by frequency-doubling
an infrared laser in KNbO$_3$, and has been described previously
\cite{ludlow01}. A diode laser master-oscillator-power-amplifier
(MOPA) system delivers 300 mW single frequency at 846 nm, as
shown in Fig. \ref{fig:exptSketch}.  This laser is phase-locked to
a build-up cavity using the Pound-Drever-Hall technique
\cite{pound83}, giving a power enhancement of 30.  A 10mm long
a-cut KNbO$_3$ crystal in the small waist of the build-up cavity
is used to generate 45 mW output power at 423 nm via non-critical
phase matching at a temperature of
$-12^{\footnotesize{\mbox{o}}}$ C \cite{ludlow01}.

The laser is further stabilized by locking the 423 nm light to the
calcium resonance transition using saturated absorption
spectroscopy in a calcium vapor cell \cite{libbrecht95}. An
acousto-optic modulator (AOM) in one arm of the saturated
absorption laser beams shifts the laser frequency so that the
laser beam sent to the MOT is 35 MHz (one natural linewidth) below
the atomic resonance.  We also use the AOM to chop this beam and
use a lock-in amplifier to eliminate the Doppler background in the
saturated absorption signal.  Because the 846 nm laser is already
locked to the frequency-doubling cavity, the feedback from this
second lock circuit servos the frequency-doubling cavity length.

\begin{figure}
\includegraphics[angle=270,width=2.6in]{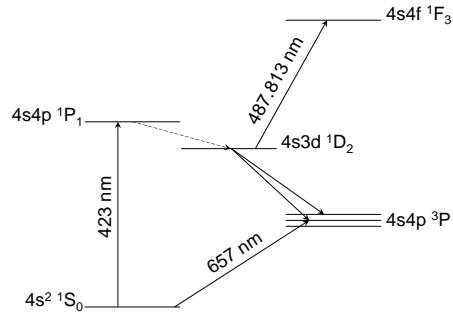}
\caption{\label{fig:levels1} A partial energy-level diagram for
calcium (not to scale).}
\end{figure}

The trap is loaded from a thermal beam of calcium atoms that
passes through the center of the MOT.  The thermal beam is formed
by heating calcium in a stainless steel oven to
$650^{\footnotesize{\mbox{o}}}$ C. The beam is weakly collimated
by the 1mm diameter, 10mm long aperture in the oven wall.  As the
beam passes through the MOT, the slowest atoms in the velocity
distribution are cooled and trapped. An additional red-detuned
(140 MHz, or four times the natural linewidth) laser beam
counter-propagates the calcium atomic beam, significantly
enhancing the MOT's capture efficiency. The density profile of the
MOT is approximately Gaussian, with a $1/e^2$-radius of 0.5 mm and
a peak density of $10^{9}$ cm$^{-3}$. The lifetime of the MOT is
limited by optical pumping to the $4s3d\;^1D_2$ state (see Fig.
\ref{fig:levels1}).

\section{Optical dipole trap}

The dipole trap is formed by focusing a 488 nm argon-ion laser
beam in the center of the MOT.  The interaction of the laser beam
with the atoms is easily described in terms of the AC-Stark shift.
The electric field of the laser beam $\vec{E}$ induces a
polarization $\vec{P}$ in the atom. The interaction of these two
fields gives rise to the $- \vec{P} \cdot \vec{E}$ potential.  A
rotating wave approximation of this interaction leads to the
well-known optical potential or ``light shift'':

\begin{equation}
U=\frac{\hbar \gamma^2}{8 \Delta}\frac{I(r)}{I_s},
\label{eqn:lightshift}
\end{equation}

\noindent where $\hbar$ is Planck's constant divided by 2$\pi$,
$\gamma=\tau^{-1}$ is 2$\pi$ times the natural line width, and
$I_s=\pi h c \gamma / 3 \lambda^3 $ is the saturation intensity,
$c$ is the speed of light, $\lambda$ is the wavelength of the
atomic transition, $\Delta = \omega - \omega_0$ is the detuning of
the laser $\omega$ from the atomic transition $\omega_0$ in rad/s,
and $I(r)$ is the intensity of the laser beam. For multi-level
atoms with many transitions from a given state, the light shift
calculation in Eq. \ref{eqn:lightshift} is extended by summing
contributions from all of the transitions connected to the level,
replacing $\gamma$ with the appropriate Einstein-A coefficients.
We use the data tabulated in Refs. \cite{nist,kurucz} .

The 488 nm argon-ion laser wave\-length is near-resonant with the
$4s3d\;^1D_2 - 4s4f\;^1F_3$ transition ($\Delta = 2\pi c (
1/487.9863\mbox{nm} -  1/487.8126\mbox{nm} ) = -1.38 \times
10^{12}$).  For a 1 W laser beam focused to a 20$\mu$m Gaussian
waist, the optical potential depth is $U/k_B = 11.6$ mK.  Such a
deep potential is required to trap our relatively hot calcium
atoms. The atomic temperature is near the Doppler limit of laser
cooling using the 423 nm transition.

Dipole traps for heavier group II atoms have been reported. Those
experiments cooled on the intercombination lines, which have a
much lower Doppler limit.  This could be done in calcium,
especially in conjunction with quenched cooling
\cite{binnewies01,curtis03} or two-photon cooling \cite{filho03}.
However, no reports have been published to our knowledge.

\section{Dipole trap loading}

We load the dipole trap while the MOT light is on.  The dipole
trap fills up with atoms optically pumped into the $^1D_2$ state
in the region of the 488 nm laser beam focus.  Other $^1D_2$ atoms
from outside the focal region pass through the dipole trap, but
there is no dissipative cooling mechanism to capture them.

Some of the $^1D_2$ atoms are ionized by the 488 nm laser beam via
a two-photon transition to the continuum. This photo-ionization
pathway is enhanced by a near-resonance with the $^1F_3$ level. As
discussed below, it is probably further enhanced by a
near-resonance in the Rydberg series leading up to the Ca {\sc II}
$3d$ ionization limit.

\begin{figure}
\includegraphics[angle=270,width=3.4in]{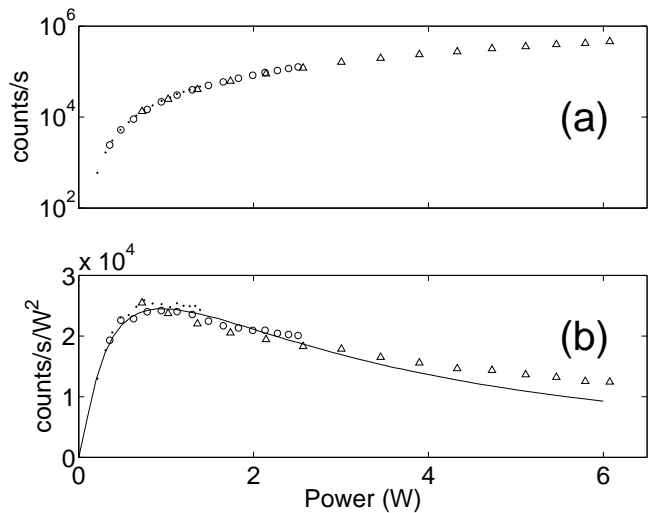}
\caption{\label{fig:trapload} Ion count rate versus 488 nm laser
power.  The top panel (a) plots the raw count rate for a
(measured) Gaussian waist of $w=20\: \mu$m.  The laser power was
changed by varying the argon-ion laser tube current.  The
different symbols show measurements made with different
intracavity laser apertures. The lower panel (b) plots the ion
count rate divided by the square of the laser power, which is
proportional to the number of $^1D_2$ atoms in the dipole trap.
The solid line is a fit of our Monte-Carlo simulation of trap
loading to the data.}
\end{figure}

We measure the ion production rate for different 488 nm laser beam
intensities \cite{channeltron}. Sample data is plotted in Fig.
\ref{fig:trapload}(a).  This data is re-plotted in Fig.
\ref{fig:trapload}(b) with the ion production rate divided by the
laser power squared.  This ratio is proportional to the number of
$^1D_2$ atoms in the laser focus, and in the absence of a dipole
trap, this signal should be a flat line. At low powers, the number
of atoms in the dipole trap increases as the trap depth increases.
The trap number maximizes when the average well depth is a few
times the Doppler temperature.  At higher power, the trap number
fall off because the 488 nm laser beam shifts the $4s^2 \; ^1S_0$
and $4s4p \; ^1P_1$ levels out of resonance with the MOT laser,
reducing the efficiency of the optical pumping loading mechanism.
We performed a Monte-Carlo simulation of trap loading and found
agreement with our data, as plotted in Fig. \ref{fig:trapload}b.

At higher laser powers, the ion production rate is somewhat higher
than expected.  This occurs when the light-shift of the $^1S_0$
and $^1P_1$ levels due to the 488 nm laser beam exceeds the
natural linewidth.  This is precisely the condition under which
ground-state atoms can be captured in the dipole trap.  For these
atoms the trap depth is comparable to the atom temperature (the
Doppler cooling limit).  Such an arrangement would increase the
density of ground-state atoms, making the loading rate due to
optical pumping higher at higher powers.

We can estimate the number of $^1D_2$ atoms in our dipole trap
using a simple rate equation.  For deep optical potentials in
steady state conditions, this is equal to the trap loading rate
multiplied by the trap lifetime. The loading rate is equal to the
optical pumping rate multiplied by the dipole trap volume, and
divided by the MOT volume.  Because the confocal parameter of the
488 nm laser beam exceeds the MOT dimension, and because the
dipole trap oscillation period along the symmetry axis is long
compared to the $^1D_2$ lifetime, we can assume that the volume
ratio is just the square of the laser beam waist divided by the
square of the Gaussian size of the MOT cloud.  The number of
$^1D_2$ atoms in the dipole trap, $N_{D}$, can be written as

\begin{equation}
N_{D} = \frac{s/2}{1 + s + (2\Delta/\gamma)^2} \;
\tau_{\mbox{\footnotesize{eff}}} A N_{S} \left(\frac{w^2}{2
r_0^2}\right) , \label{eqn:steadystate}
\end{equation}

\noindent where $s=I/I_s$ is the saturation parameter, $N_S$ is
the number of ground state atoms in the trap,
$\tau_{\mbox{\footnotesize{eff}}}$ is the dipole trap lifetime,
$A=2150$ s$^{-1}$ is the Einstein A coefficient for the $^1P_1 -
^1D_2$ transition, and $r_0$ is the Gaussian $1/e^2$ radius of the
MOT. Depending on the beam waist and laser power, this simple
model tells us that we load up to 2000 atoms into the trap, for a
peak density of approximately $5\times 10^8$ cm$^{-3}$.  By
comparison, the background density of $^1D_2$ atoms is given by
Eq. \ref{eqn:steadystate} without the volume ratio, and with the
numbers $N_D$ and $N_S$ replaced by densities. The $^1D_2$
background density is $\sim 10^7$ cm$^{-3}$.

In view of these numbers, it is perhaps surprising that at the
lowest laser powers we can detect a small number of atoms in the
dipole trap, and discriminate against the background $^1D_2$
atoms. But the background atoms roll through the trap in a few
$\mu$s, and the trapped atoms remain in the trap approximately 100
times longer. The photo-ionization probability increases with the
time spent in the 488 nm laser focus, so the ionization signal is
predominantly from the trapped atoms.

\section{Two-photon photoionization rate}

We measure the lifetime of the dipole trap by blocking the MOT
laser beams, and measuring the decay of the ion signal. The three
most important decay mechanisms are radiative decay of the $^1D_2$
level, collisional decay due to hot atoms from the thermal atomic
beam, and two-photon ionization of the $^1D_2$ atoms. Because we
do not have a suitable method for turning off the thermal beam, we
cannot reliably extract the $^1D_2$ radiative lifetime.  Published
values of the lifetime are around 2 ms
\cite{pasternack80,husain86,drozdowski93,porsev01,fischer03},
somewhat longer than measured in our experiment.  While our
experiment cannot determine the radiative lifetime, by measuring
the decay rate as a function of 488 nm laser intensity we can
determine the photo-ionization cross-section.

A rate-equation for $^1D_2$ level decay in the dipole trap after
the loading has turned off is

\begin{equation}
\frac{dN_{D}}{dt} = -N_{D}\left(
\frac{1}{\tau_{\mbox{\footnotesize{eff}}}} + {\cal A}I^2 \right),
\label{eqn:rate}
\end{equation}

\noindent where ${\cal A}$ is the two-photon ionization rate
coefficient. This has the well-known solution

\begin{equation}
N_D(t) = N_D(0)
\exp\left(\frac{1}{\tau_{\mbox{\footnotesize{eff}}}}+{\cal
A}I^2\right).
\end{equation}

In second-order perturbation theory, two-photon ionization is
written as an overlap of the initial and final states summed over
all possible intermediate states, divided by an energy
denominator.  For near-resonant ionization, the energy denominator
makes the near-resonant term dominant, collapsing the sum to just
one term.  This one term looks like the product of the probability
that an atom is excited into the $^1F_3$ state multiplied by the
probability of photo-ionizing out of that state.  We can write
this term as

\begin{eqnarray}
{\cal A} I^2 & = & \frac{s/2}{1+s+(2\Delta/\gamma)^2}
\frac{I}{h\nu}
\sigma \\
 & = & \frac{3 \lambda^4 \sigma \gamma }{8 \pi h^2 c^2
 \Delta^2} I^2 \label{eqn:aisquared},
\end{eqnarray}

\noindent where $\sigma$ is the $^1F_3$ one-photon
photo-ionization cross section. The approximation in Eq.
\ref{eqn:aisquared} assumes that $s << (2\pi\Delta/\gamma)^2$ and
$\gamma << \Delta$.  In our experiment, the intensity of the 488
nm laser has a Gaussian spatial profile.  Averaging the square of
the intensity over the laser profile allows us to relate the
photo-ionization rate to the total laser power.  In this case, it
can be written as

\begin{equation}
{\cal A}I^2 = \frac{\lambda^4 \sigma \gamma}{2 \pi^3 h^2 c^2
\Delta^2 w^4}P^2. \label{eqn:powerlaw}
\end{equation}

\begin{figure}
\includegraphics[width=3.4in]{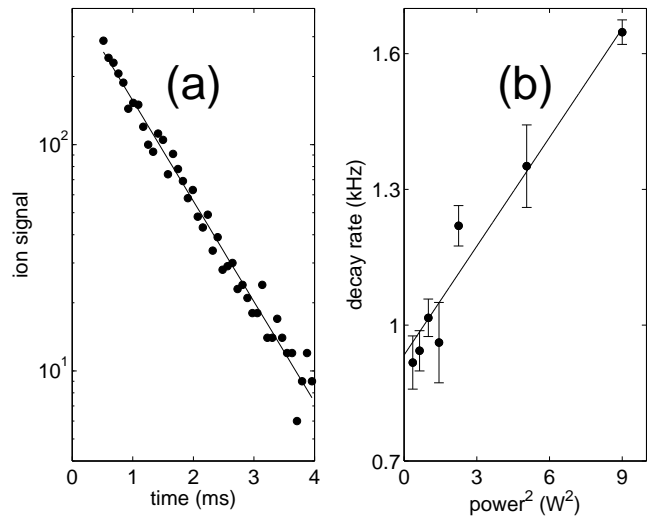}
\caption{\label{fig:decay} Dipole trap decay versus power squared.
The decay signal from the dipole trap.  The left panel (a) shows
the ion signal versus time with 1 W focused to a 90 $\mu$m waist.
The effective decay rate for this data is 1.02 kHz.  The right
panel (b) shows the apparent decay rate for a range of powers, all
focused to a 90 $\mu$m waist.  The decay rate is fit to a line in
square of the power, as suggested by Eq. \ref{eqn:rate}. }
\end{figure}

The dipole trap decay rate as a function of the square of the 488
nm laser power squared is shown in Fig. \ref{fig:decay}. For each
power level, we measured the dipole trap decay.  At sufficiently
low power levels, this decay is approximately exponential, and we
extract the decay rate using a least-squares fitting routine
\cite{i-squared}. The decay rate depends on power.  The
zero-extrapolated decay rate is
$\tau_{\mbox{\footnotesize{eff}}}^{-1} = 0.93$ kHz. This rate is
approximately twice the radiative decay rate.  We see evidence in
our experiment that our rate is significantly influenced by
collisions with atoms in the thermal atomic beam.  This is not
surprising because the dipole trap sits in the middle of the
thermal atomic beam.  Without detailed characterization of the
thermal beam, we cannot reliably extract a
thermal-atom-$1D_2$-atom collision cross-section.

The slope of the decay rate with power is $80$ Hz/W$^2$.  Using
Eq. \ref{eqn:powerlaw}, we can determine the $^1F_3$
photo-ionization cross section.  This gives a photo-ionization
cross-section for the $^1F_3$ level of $\sigma = 230 \times
10^{-18}$ cm$^2$.  This extraordinarily large cross section
suggests that the final state lies near a Rydberg state in the
continuum.  The final state is 133 cm$^{-1}$ below the Ca {\sc II}
$^2D_{3/2}$ ionization limit. The principle quantum of hydrogenic
Rydberg levels in this region are $n\sim 29$, and the separation
between levels is 9 cm$^{-1}$.  Our measurements are carried out
in the presence of an electric field, further increasing the
probability of finding a nearby Rydberg level.  The NIST database
tabulates only odd parity levels in this energy region. We are
unaware of applicable quantum defect calculations or measurements.

\section{Conclusion}

We have demonstrated an optical dipole trap for neutral calcium
atoms. These atoms are non-adiabatically loaded into the trap by
an optical pumping mechanism.  The lifetime of our trap is limited
by the $\sim 2$ ms lifetime of the $^1D_2$ atoms and by collisions
with atoms in the thermal atomic beam.

Our initial interest in the calcium optical dipole trap was its
potential application to our ultracold plasma research. It may be
possible to use the dipole trap as the beginning point for
ultracold plasma expansion studies.  Because of the high aspect
ratio, a plasma generated from a dipole trap would be a
two-dimensional ultracold neutral plasma at early times.
Correlation heating would be reduced compared to the
three-dimensional case \cite{chen04}, bringing the two-dimensional
plasma closer to the strongly-coupled regime.  Furthermore, the
plasma expansion depends on the exact density distribution of the
initial cloud.  Compared to standard MOT traps, a dipole trap has
a well-defined density profile, meaning that these new plasmas
could greatly improve the reproducibility of plasma expansion
experiments.  We plan to explore these possibilities in future
work.

\section{Acknowledgements}

This work was supported in part by grants from the Research
Corporation and the National Science Foundation (Grant No.
PHY-9985027).


\begin{thebibliography}{99}
\bibitem{channeltron}
A weak elec\-tric field ($\sim$5 V/cm) ap\-plied to
high-transparency grids on either side of the MOT extracts the
ions. The ions are detected using a channeltron (Burle model 4860)
and counted in a multichannel scaler.
\bibitem{i-squared}
The trap depth is proportional to $I$, and the photoionization
signal is proportional to $I^2$.  At high laser powers, when atoms
can be trapped even at large radii, it is possible that kinetic
effects will skew the photo-ionization signal.  Atoms trapped at
the bottom of the potential will be ionized first, and atoms
trapped at the edge of the potential will be ionized at a later
time.  For low enough laser powers, when the potential is too
shallow to trap significant numbers of atoms, this effect should
not influence the photo-ionization signal.
\end{thebibliography}
\end{document}